\documentstyle[12pt,aaspp4]{article}
\newcommand{\ifpp}[1]{#1}
\newcommand{\ifms}[1]{}
\ifpp{\input psfig}
\newcommand{\degree}{\ifmmode {^{\circ}} \else {$^{\circ}$} \fi}
\newcommand{\degrees}{\ifmmode {^{\circ}} \else {$^{\circ}$} \fi}

\newcommand{\unit}[1]{\ifmmode {\rm\ #1\,} \else {$\rm #1$} \fi}
\newcommand{\quarter}{\ifmmode {\frac{1}{4}} \else {$\frac{1}{4}$} \fi}
\newcommand{\angstrom}{\unit{\AA}}
\newcommand{\angstroms}{\unit{\AA}}

\newcommand{\etal}{{\em et al.~}}

\newcommand{\tten}[1]{\ifmmode {\times 10^{#1}} \else {$\times 10^{#1}$} \fi}
\newcommand{\tentothe}[1]{\ifmmode {10^{#1}} \else {$10^{#1}$} \fi}

\newcommand{\doublet}{\ifmmode {\lambda\lambda} \else {$\lambda\lambda$} \fi}
\newcommand{\singlet}{\ifmmode {\lambda} \else {$\lambda$} \fi}

\begin{document}

\lefthead{Korpela \& Bowyer}
\righthead{EUVE Observations of Neutron Stars}

\title{EUVE Observations of Neutron Stars}
\author{Eric J. Korpela and Stuart Bowyer}
\affil{Space Sciences Laboratory, University of California, Berkeley, CA 94720}

\begin{abstract}
We present the results of searches for EUV emission from neutron stars 
conducted with the EUVE Deep Survey and Scanner Telescopes.  To date, 21 fields
containing known neutron stars have been observed in the Lexan/Boron 
(40--190 \angstrom) band. Of these,
11 fields were simultaneously observed in the Aluminum/Carbon 
(160-385 \angstrom) band.
Five neutron stars which have been detected in the EUV have been reported
previously; no new detections have been made in the studies reported here.
For those sources not
detected, we have used the observations to obtain limits on the spectral flux from the neutron stars in these bands.
We provide means to convert these fluxes into intrinsic source fluxes
for black-body and power law spectra for varying levels of absorption by
the interstellar medium.
\end{abstract}


\section{Introduction}

The study of EUV emission from neutron stars can address several problems
in the astrophysics of neutron stars.  Thermal emission from the star can be 
compared with models of the cooling of neutron star interiors. As 
processes in the neutron star atmosphere may be important in modifying 
this thermal radiation, studies of EUV emission can lead to clues to the
composition and structure of neutron star atmospheres (Pavlov \etal, 1996).
For older neutron stars, the possibility of reheating due to
internal and external processes can be investigated.(Edelstein \etal, 1995).
In addition magnetospheric processes can contribute to EUV emission, providing
constraints on the magnetic field and magnetospheric plasma properties.

Due to interstellar absorption, EUV observations of neutron stars are
difficult.  The earliest EUV observations of radio pulsars were made by
Greenstein \etal (1977) 
using an EUV telescope carried by the Apollo-Soyuz mission.  Their observation
produced only upper limits to the EUV flux of $\sim 10^{-12} {\rm erg/s/cm^{2}/ \angstrom}$ for three sources.  
Observations at soft X-ray bands using Einstein and ROSAT have 
produced detections of several neutron stars (\"{O}gelman, 1995).
The Extreme Ultraviolet Wide Field Camera (WFC) on ROSAT did not report
any detection of EUV emission from neutron stars.

A number of neutron stars have been detected in the extreme 
ultraviolet with the Extreme Ultraviolet Explorer (EUVE) satellite.  One
of these is in an accretion system (Her X-1) where the bulk of the emission
is due to the accretion disc; others are neutron stars 
radiating thermally or with a nonthermal pulse related emission. 
All of the detections were made in connection with 
long duration Guest Observer studies of neutron stars. A number of searches
for EUV emission from neutron stars have been carried out.  Some of these were
short duration exploratory searches of
likely candidates carried out by the authors as Guest Observers.
Some fields were examined in the EUVE ``Right Angle'' observation
program.

In this paper we provide the results of all the EUVE observations of
neutron stars.
Twenty fields containing  
known neutron stars have been observed in the Lexan/Boron (40--190 \angstrom) band. 
Of these,
11 were simultaneously observed in the Aluminum/Carbon (160-385 \angstrom) band.
We include the previously reported detections for 
the purpose of providing a complete sample of the
sources observed by EUVE.
For those sources not
detected, we have used the observations to obtain limits on the spectral flux in
the Lexan/Boron and Aluminum/Carbon bands.
We present these results and provide means to convert these fluxes into intrinsic source fluxes 
for black-body and power law spectra, and for varying levels of absorption by
the interstellar medium.

\section{Observations}
The EUVE satellite (Bowyer \& Malina 1991) offers two observing modes for broadband filter photometry.  The
primary instrument, the ``Deep Survey'' telescope, has a high effective area and provides filters
for both the 100 \angstrom Lexan/Boron band and the 200 \angstrom Aluminum/Carbon band.  These filters cover different portions of the field of view. Therefore
a target cannot be simultaneously observed in both bands with the Deep Survey
telescope.
As the primary science instrument, this telescope is usually used for 
Guest Observer projects.  Some Guest Observer time has been granted to 
the authors for observations of fields containing
neutron stars; we include the results of those observations here.

The EUVE satellite also has 3 ``Scanner'' telescopes set at right angles to the
Deep Survey telescope.  The three coaxial telescopes have smaller effective areas, but
allow simultaneous observation in both the 100 \angstrom and 200 \angstrom bands.  Because the satellite
can rotate around the axis of the Deep Survey telescope,
these telescopes may be used when a potentially interesting object
falls $90\pm 2\degrees$ from a target
being observed with the Deep Survey telescope.  
We (and others) have provided lists of neutron stars for possible observation
using this method.  Although there is a low
probability of any specific target being observed using this approach, 
a reasonable number of fields have been observed.
Because the exposures are usually of the 
same duration as that of the Deep Survey observation, these observations
can be quite long. 

The bandpasses in the Deep Survey and Scanner telescopes are defined by thin 
film filters.
The Lexan/Boron 100 \angstrom band extends from 40--190 \angstroms and is sharply peaked at
90 \angstroms.  The Aluminum/Carbon 200 \angstrom band extends from 160--385 \angstroms and
is peaked at 170 angstroms.  The effective areas of these filters are provided
in Bowyer \etal (1996). Because the variation of the absorption cross section 
of the interstellar medium with wavelength is substantial in the EUV, differing
interstellar absorptions can greatly alter the effective bandpass.  This effect will be
discussed in Section 3.

The data discussed herein were obtained through the Center for EUV Astrophysics,
(CEA) which operates the EUVE satellite.  The data were processed through the
CEA data pipeline which takes raw satellite data, screens out bad data due
to high noise rates or SAA passages, corrects for instrument distortions,
and calculates effective exposure times.  The output products of this pipeline 
are images in sky coordinates.

The point spread function (PSF) of the EUVE instruments is $\sim 30^{\prime\prime}$ FWHM.  This
is comparable to the RMS EUVE position error for identified sources of $\sim 45^{\prime\prime}$.  Because of
these uncertainties, and the fact that the positions of some neutron stars are uncertain by up to several arcminutes,
we performed an exhaustive search for point sources within $\sqrt{1+\sigma_{\rm psr}^2}$ arcminutes
of the expected neutron star position in the images, where $\sigma_{\rm psr}$ is the uncertainty in the
neutron star position.  In those cases where no sources were found, we present
an upper limit which is $2 \sigma$ above the maximum PSF convolved count rate found in the search area. The values for $\sigma$ are calculated through count statistics using the relation
\begin{equation}
\sigma=\sqrt{N_{fg}+ N_{bg} \frac{A^2_{psf}}{A^2_{bg}}}
\end{equation}
where
$N_{fg}$ indicates the number of counts in a PSF convolved bin of given 
coordinates and
$N_{bg}$ is the number of counts in a background annulus of
inner radius 1$^\prime$ and outer radius 3$^\prime$ centered on the same coordinates. 
$A_{psf}$ is the effective solid angle of the autoconvolved PSF ($\sim 1.2$ arcmin$^2$), and $A_{bg}$ is the solid angle of the background annulus ($\sim 25$ arcmin$^2$).
In cases where an object was located near the edge of the detector or in a 
region
of rapidly varying background, the size and shape of the background region
was modified, either by moving the background annulus to an unaffected region,
or changing the inner and outer radii of the annulus. The error calculations
were modified accordingly to consider the solid angle of the new background
region.
The count rate measurements or upper limits were corrected for instrument
vignetting and instrument dead-time and converted to a 
mean spectral flux over the measurement bandpass by dividing by the integrated effective area (\cite{edelstein,euvearea})
and multiplying by the bandpass mean photon energy.  The results are shown
in Table~1, accompanied by relevant data on the neutron star obtained from a variety
of sources.  For those sources which had been previously detected, we provide
the previously measured count rate and a calculation of flux using the method
described above.

\ifpp{
\begin{table}[p]
\begin{center}
\plotone{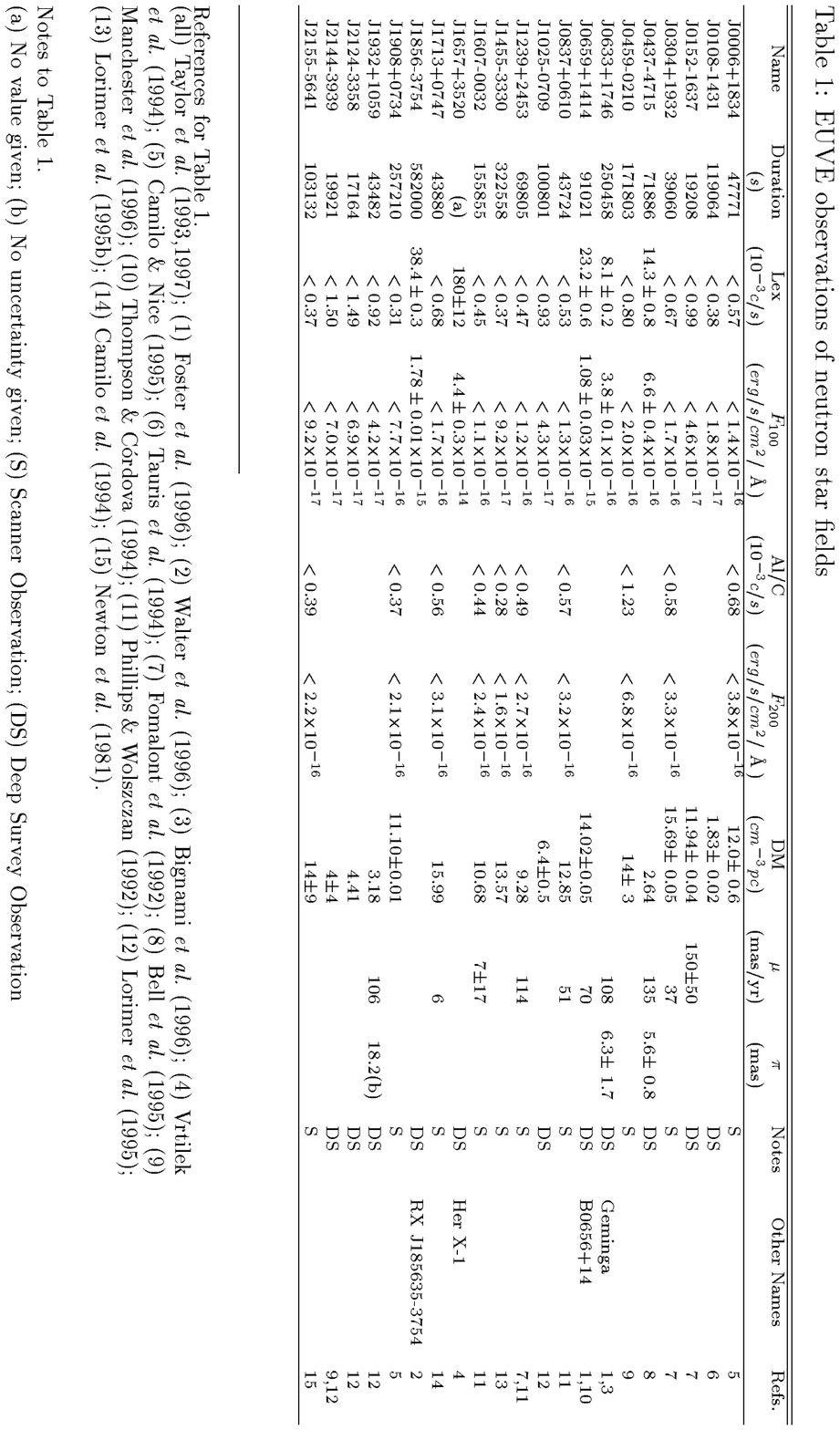}
\end{center}
\end{table}
}

\section{Discussion}

The limits obtained here may or may not be important in constraining models
of pulsar structure and emission mechanisms.  The importance depends upon
the emission model, the distance to the pulsar, and the interstellar 
absorption in the direction of the pulsar.  Because these values are,
in general, unknown, we provide factors to convert the measured EUV flux
into an intrinsic source flux over a range of values.  

Because the opacity is large in the EUV and varies rapidly with
wavelength ($\kappa \stackrel{\propto}{\sim} \lambda^3$), an approximation of constant opacity over the
bandpass is not applicable. Increasing absorption by the interstellar medium has the effect of reducing the effective area
of the EUV telescopes at longer wavelengths.  This shifts the peak and mean wavelengths
of the effective bandpass to shorter wavelengths.  Because of this effect,
the intrinsic spectral shape of the source has an impact upon how greatly the source is 
affected by the absorption.  For example, a $10^5$ K
black-body is more affected than a $10^6$ K black-body 
because it emits more energy at longer wavelengths.    
Unfortunately, there is no
good analytical approximation for these effects.  Using the interstellar opacity cross sections of
Rumph \etal (1994), we have calculated the effects of interstellar absorption
on the measured count rate for a variety of hydrogen column densities 
for both black-body and power law spectra.  We have
incorporated these results in
conversion factors (discussed below) which allow one to
convert the measured flux to an intrinsic flux for a
range of black-body temperatures and power law indices.

In Figure~1 we show the conversion factor $C_{100}^{bb}$ required to convert the measured 100 \angstrom flux to
an intrinsic unabsorbed flux at 100 \angstroms for a black-body at a given temperature, using the relationship $F_{\rm o}(100 \angstroms)
= C^{bb}_{100} F_{100}$.  We obtained this conversion factor by calculating
an absorbed black-body spectrum and determining the associated
Lexan band count rate.  This count rate was converted to a band averaged
flux by dividing by the integrated effective area of the Lexan band.  The
ratio of the unabsorbed flux to this averaged flux is the conversion factor. 
The conversion factor is plotted for neutral hydrogen column densities of 
$0, 10^{19}, 10^{19.5}, 10^{20},$ and $10^{20.5}$. Note that these
conversion factors can be less than unity in those cases where the emission
is predominantly at wavelengths longer than the bandpass central wavelength.
In Figure~2 we show the equivalent conversion factors
for the 200 \angstrom band.

\ifpp{
\begin{figure}
\begin{center}
\ \psfig{file=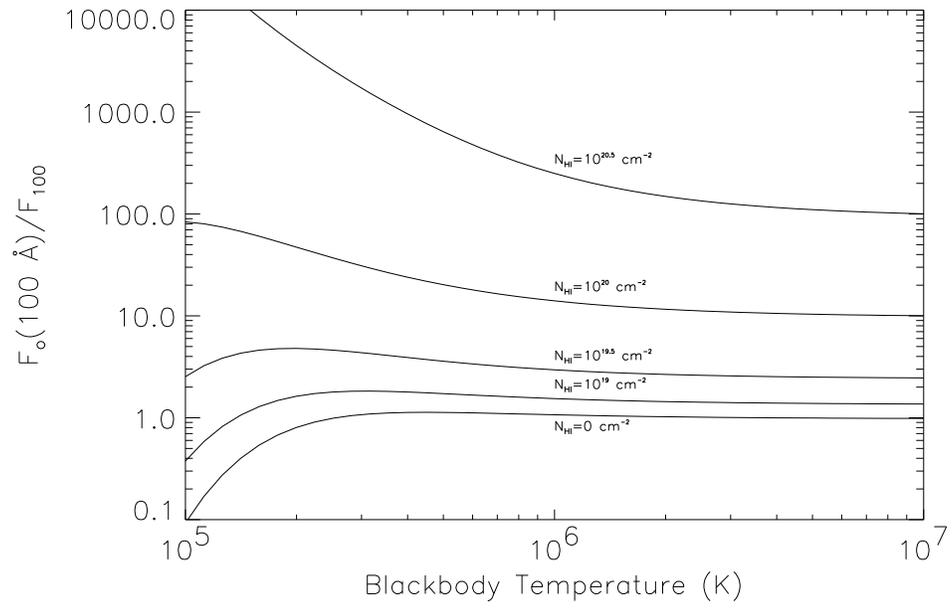}
\caption{Lexan/Boron band absorption conversion factors for black-body spectra.} 
\end{center}
\end{figure}
\begin{figure}
\begin{center}
\ \psfig{file=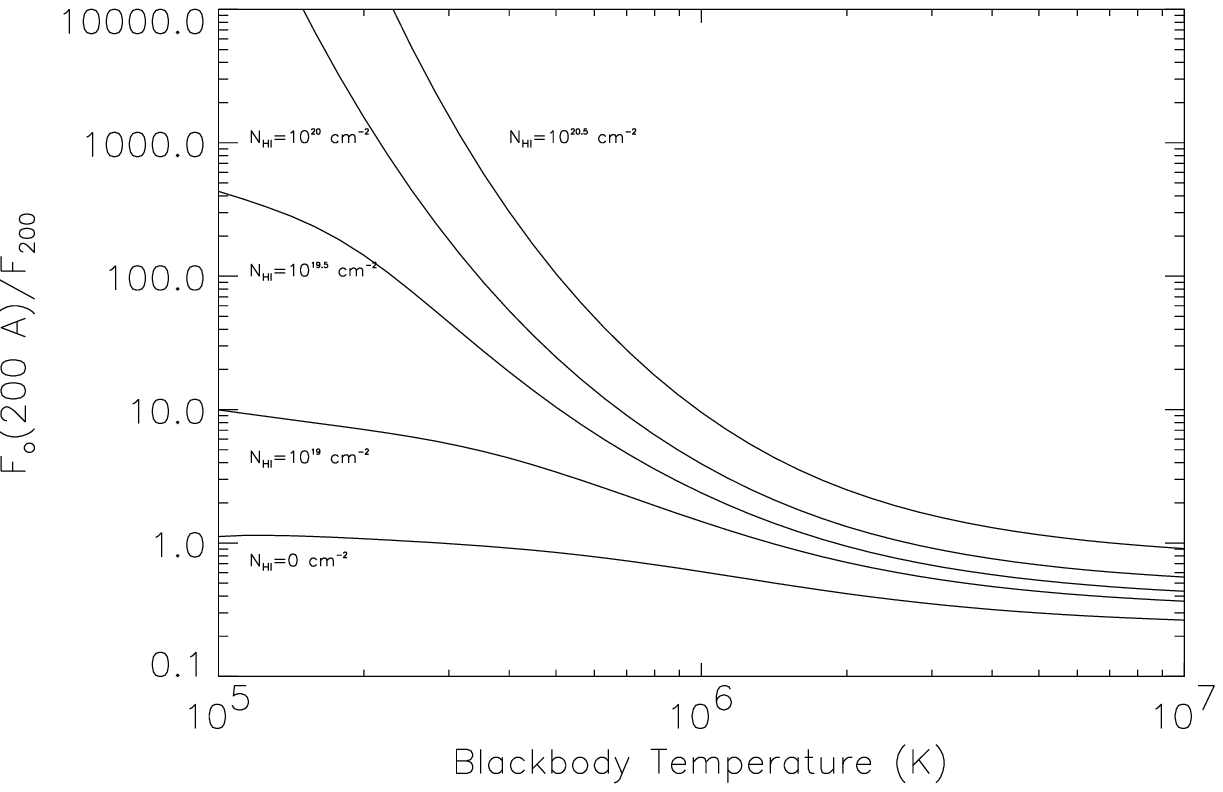}
\caption{Aluminum/Carbon band absorption conversion factors for black-body spectra.} 
\end{center}
\end{figure}
}

In Figure~3 we show the 100 \angstrom band conversion factors $C_{100}^{pl}$ for power law spectra of the form 
$F_{\lambda}=F_{o} \left( \frac{\lambda}{\lambda_{o}} \right)^{\alpha}$ for power law indices
between -4 and 4.  
In Figure~4 we
show the equivalent conversion factors for the 200 \angstrom band.

\ifpp{
\begin{figure}
\begin{center}
\ \psfig{file=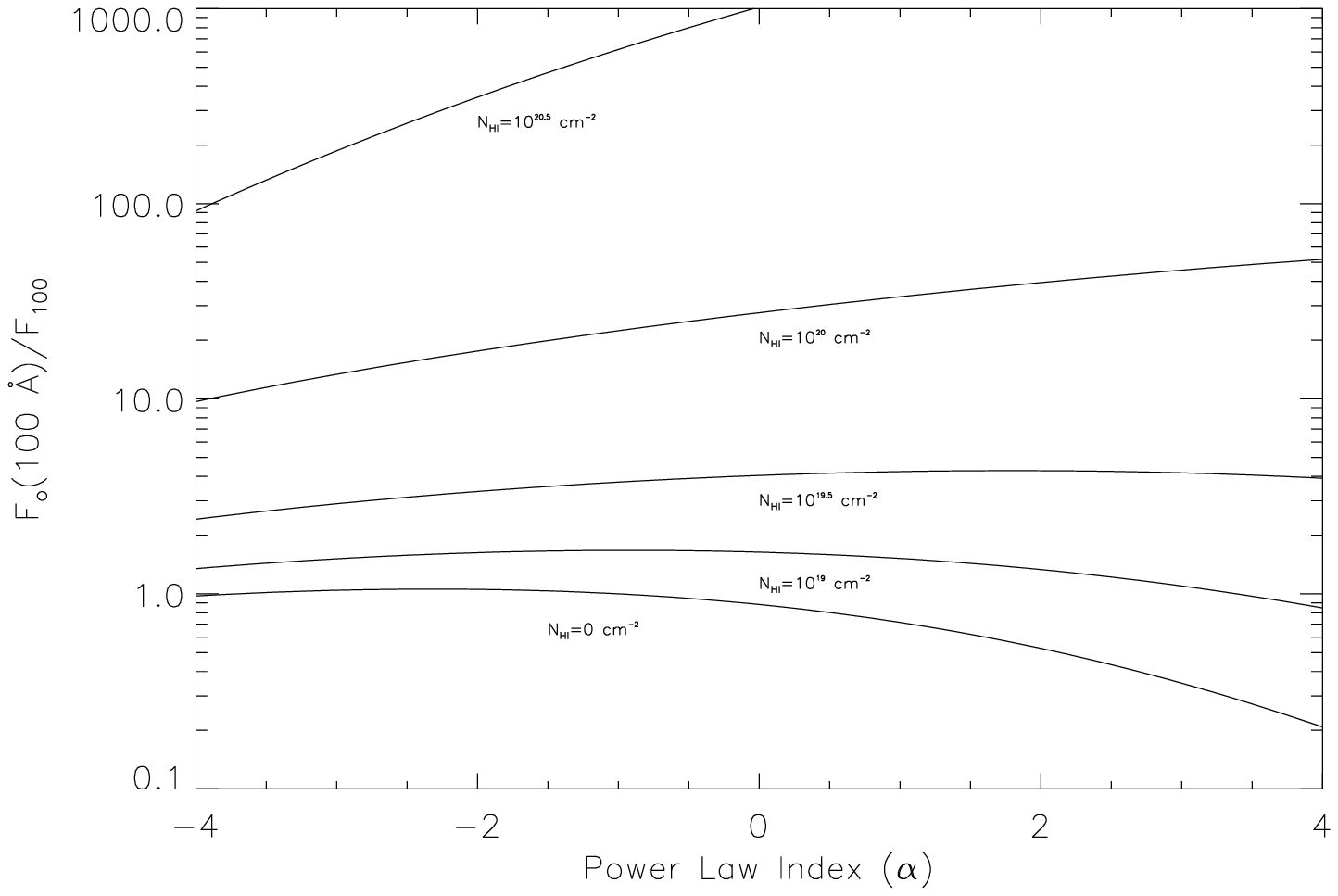}
\caption{Lexan/Boron band absorption conversion factors for power law spectra.} 
\end{center}
\end{figure}
\begin{figure}
\begin{center}
\ \psfig{file=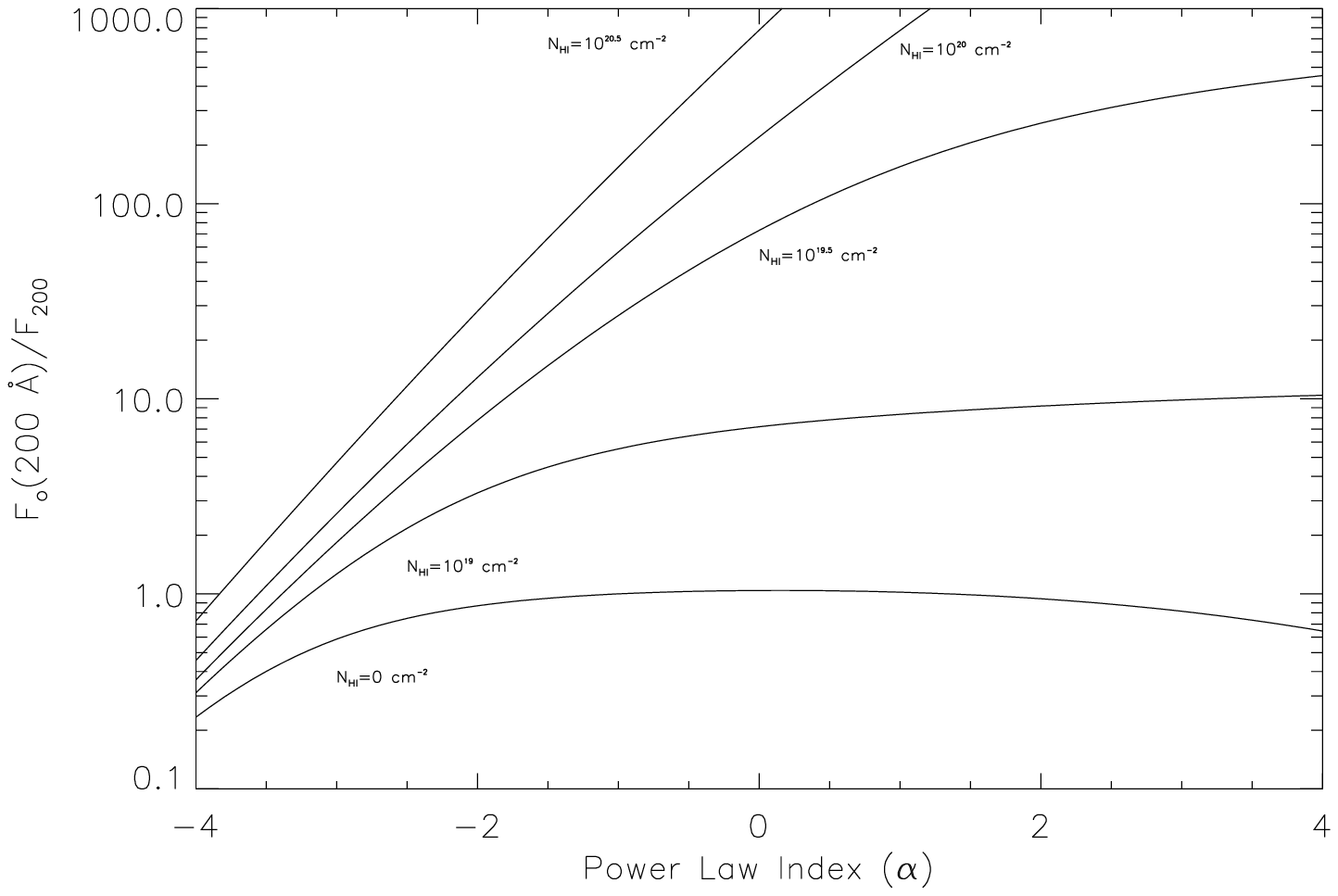}
\caption{Aluminum/Carbon band absorption conversion factors for power law spectra.} 
\end{center}
\end{figure}
}

\section{Summary}
We present the results of searches for EUV emission from 21 neutron stars.
Five neutron stars (reported previously) have been detected with EUVE 
in the 100 
\angstrom Lexan/Boron band.  
We provide
measured fluxes for these sources and upper limit fluxes for the remaining 
objects which have not been 
detected.  We also provide upper limit fluxes in the 200 \angstrom Aluminum/Carbon
band for those sources which were observed in that band.   Of the five detected neutron stars,
one is an active accretion source (Her X-1), one is a millisecond pulsar 
(J0437-4715), two are isolated pulsars (Geminga, J0659+1414), and one is
an thought to be an isolated, non-pulsing neutron star (RX J185635-3854).

\ \\
We thank Michael Lampton for useful comments.
This work has been supported by NASA grant NAS5-30180.

\ifms{
\clearpage
\figcaption[korpela.fig1.eps]{Lexan/Boron band flux conversion factors for black-body spectra.} 
\figcaption[korpela.fig2.eps]{Aluminum/Carbon band flux conversion factors for black-body spectra.} 
\figcaption[korpela.fig3.eps]{Lexan/Boron band flux conversion factors for power law spectra.} 
\figcaption[korpela.fig4.eps]{Aluminum/Carbon band flux conversion factors for power law spectra.} 
}

\end{document}